\documentclass[noshowpacs,amsmath,
twocolumn,
superscriptaddress,
8pt,
aps,prb]{revtex4-1}
\usepackage{setspace}
\usepackage{amsmath}
\usepackage{graphicx}
\usepackage[nearskip,margin = 0pt]{subfig}

\usepackage{verbatim}
\usepackage{amsfonts}
\usepackage{amssymb}
\usepackage{epstopdf} 
\usepackage{xcolor}
\usepackage{verbatim}
\usepackage{multibib}
\DeclareGraphicsExtensions{.pdf,.eps,.png,.jpg,.mps} 
\begin{document}
\title{Single-mode dispersive waves and soliton microcomb dynamics}

\author{Xu Yi$^{1,\ast}$, Qi-Fan Yang$^{1,\ast}$, Xueyue Zhang$^{1,2,\ast}$, Ki Youl Yang$^{1}$, and Kerry Vahala$^{1,\dagger}$\\
$^{1}$T. J. Watson Laboratory of Applied Physics, California Institute of Technology, Pasadena, California 91125, USA.\\
$^{2}$Department of Microelectronics and Nanoelectronics, Tsinghua University, Beijing 100084, P. R. China\\
$^{\ast}$These authors contributed equally to this work.\\
$^{\dagger}$Corresponding author: vahala@caltech.edu}

\begin{abstract}
Dissipative Kerr solitons are self-sustaining optical wavepackets in resonators. They use the Kerr nonlinearity to both compensate dispersion and to offset optical loss. Besides providing insights into nonlinear resonator physics, they can be applied in frequency metrology, precision clocks, and spectroscopy. Like other optical solitons, the dissipative Kerr soliton can radiate power in the form of a dispersive wave through a process that is the optical analogue of Cherenkov radiation. Dispersive waves typically consist of an ensemble of optical modes. A limiting case is demonstrated in which the dispersive wave is concentrated into a single cavity mode. In this limit, its interaction with the soliton is shown to induce bistable behavior in the spectral and temporal properties of the soliton. Also, an operating point of enhanced repetition-rate stability is predicted and observed. The single-mode dispersive wave can therefore provide quiet states of soliton comb operation useful in many applications. 

\end{abstract}

\date{\today}

\maketitle

A new dissipative soliton \cite{akhmediev2008dissipative} has recently been observed in optical resonators. These dissipative Kerr solitons (DKS) have been demonstrated in fiber resonators \cite{leo2010temporal} and in various microcavity systems \cite{herr2014temporal,yi2015soliton,brasch2016photonic,wang2016intracavity,joshi2016thermally}. In microcomb research \cite{del2007optical,kippenberg2011microresonator} soliton formation produces phase-locked spectra with reproducible envelopes, as required in frequency comb applications \cite{liang2015high,del2016phase,brasch2016self,suh2016microresonator,marin2016microresonator}. Moreover, their unusual properties and interactions create a rich landscape for research in nonlinear optical phenomena \cite{jang2013ultraweak,xue2015mode,brasch2016photonic,karpov2016raman,yi2016theory,yang2016spatial,yang2016stokes,guo2016universal,bao2016observation,yu2016breather,cole2016soliton}. Two such phenomena, the Raman-induced soliton-self-frequency-shift (SSFS) and dispersive-wave generation, are important to this work. 

The Raman SSFS causes a spectral red shift of the soliton. In optical fiber systems, this shift continuously increases with propagation distance\cite{mitschke1986discovery,gordon1986theory}, however, in microresonators the shift is fixed and depends upon soliton power \cite{milian2015solitons,karpov2016raman,yi2016theory,anderson2016measurement}. Dispersive waves also occur in optical fiber systems\cite{wai1986nonlinear}. They are formed when a soliton radiates into a spectral region of normal dispersion\cite{akhmediev1995cherenkov,brasch2016photonic}. This radiation, which can also be engineered to occur through spatial mode interactions \cite{matsko2016optical,yang2016spatial}, can be understood as the optical analog of Cherenkov radiation \cite{akhmediev1995cherenkov}. Dispersive waves provide a powerful way to spectrally broaden a soliton within a microresonator as a precursor to self referencing \cite{li2015octave,brasch2016self}. Their formation also induces soliton recoil \cite{akhmediev1995cherenkov} which, similar to SSFS, causes a frequency shift in the spectral center of the soliton \cite{brasch2016photonic,karpov2016raman}. 

In microcavities, the Raman SSFS enables controlled tuning of the emission  wavelength of a dispersive wave by varying the cavity-pump detuning frequency\cite{yang2016spatial} ($\delta \omega \equiv \omega_0 - \omega_p$ where $\omega_p$ is the pump frequency and $\omega_0$ is the frequency of the cavity mode that is being pumped). This happens because the DKS repetition rate is coupled to $\delta \omega$ by the Raman SSFS\cite{yi2016theory,yang2016spatial}. As the repetition rate varies with $\delta \omega$, the phase matching condition of the soliton to the dispersive wave is also varied, thereby tuning the dispersive wave. Measurements of this coupling have been used to determine the dispersion of the soliton-forming mode family \cite{yang2016spatial}.

Dispersive waves normally consist of an ensemble of modes. However, a single cavity mode can exhibit behavior similar to that of a dispersive wave as a result of an avoided-mode crossing\cite{lucas2016study}. In this work, an avoided-mode crossing is used to excite a dispersive wave consisting of a single cavity mode. The coupling of the {\it single-mode dispersive wave} to the soliton is strongly influenced by the total soliton frequency shift produced by the combined Raman-induced SSFS and the dispersive-wave recoil. The combination is shown to induce hysteresis behavior in soliton properties. Included in this behavior, there is an operating point of improved pulse-rate stability wherein reduced coupling of repetition rate and cavity-pump detuning occurs. Pulse-rate stability is centrally important in many frequency comb applications \cite{papp2014microresonator,liang2015high,suh2016microresonator} and the fundamental contributions to phase noise in the pulse train have been considered theoretically\cite{matsko2013timing}. Technical noise mechanisms are also present. For example, DKS generation using on-chip silica resonators exhibits phase noise that tracks in spectral profile the phase noise of the optical pump \cite{yi2015soliton}. The quiet operation point is shown to reduce these noise contributions in the soliton pulse repetition rate.

In what follows both the hysteresis and the regime of improved stability (quiet point) are measured and then modeled theoretically. 

\begin{figure*}[!ht]
\captionsetup{singlelinecheck=no, justification = RaggedRight}
\includegraphics[width=18cm]{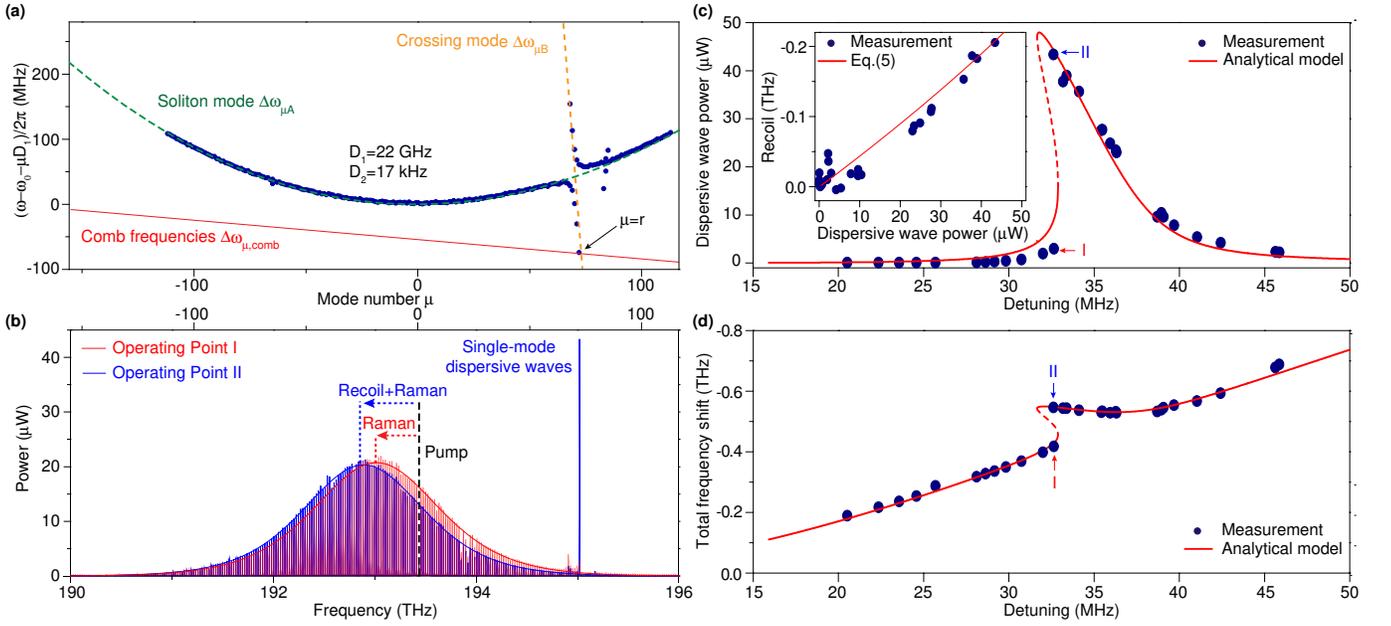}
\caption{{\bf Soliton hysteretic behavior induced by mode-interaction.} {\bf a,} Measured relative mode frequencies (blue points)\cite{yi2015soliton}. The green and yellow dashed lines represent the fitted unperturbed soliton-forming mode family A and crossing mode family B, respectively. The red line illustrates the frequencies of a hypothetical soliton frequency comb. A non-zero slope on this line arises from the repetition rate change relative to the FSR at mode $\mu=0$. {\bf b,} Measured soliton optical spectra at two operating points corresponding to closely matched cavity-pump detuning frequencies, $\delta \omega$. A strong single-mode dispersive wave at $\mu=72$ is observed for operating point II, resulting in a recoil of the soliton central frequency beyond that resulting from the Raman-induced SSFS.   {\bf c,d,} Dispersive-wave power and soliton spectral center frequency shift versus cavity-pump detuning. Inset in 1c: Measured (blue dots) and theoretical (red line) recoil frequency versus the spur power.}
\label{figure1}
\end{figure*}

\medskip

\begin{figure*}
\captionsetup{singlelinecheck=no, justification = RaggedRight}
\includegraphics[width=18cm]{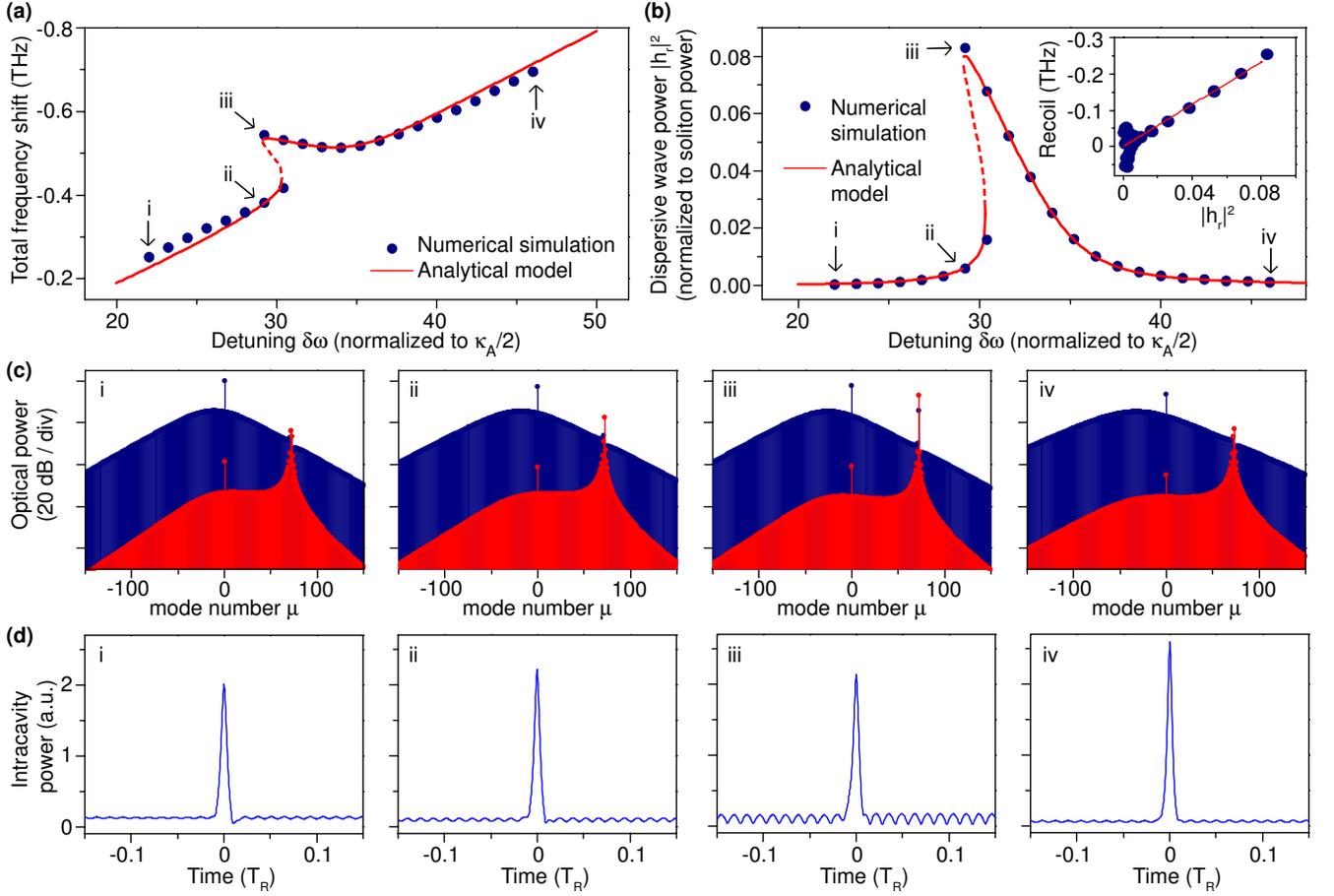}
\caption{{\bf Numerical simulation  and analytical model of single-mode dispersive wave generation and recoil.} {\bf a,} Numerical (blue dots) and analytical (red solid line) soliton total frequency shift versus cavity-pump detuning. {\bf b,} Numerical (blue dots) and analytical (red solid line) dispersive wave power (normalized to total soliton power) versus cavity-pump detuning. Inset: recoil frequency versus the dispersive wave power. {\bf c,} Optical spectra in the two mode families (blue: soliton forming mode family A; red: crossing mode family B). {\bf d,} Time domain intracavity power. $T_R$ is the cavity round trip time.}
\label{figure2}
\end{figure*}

\begin{figure*}[!ht]
\captionsetup{singlelinecheck=no, justification = RaggedRight}
\includegraphics[width=18cm]{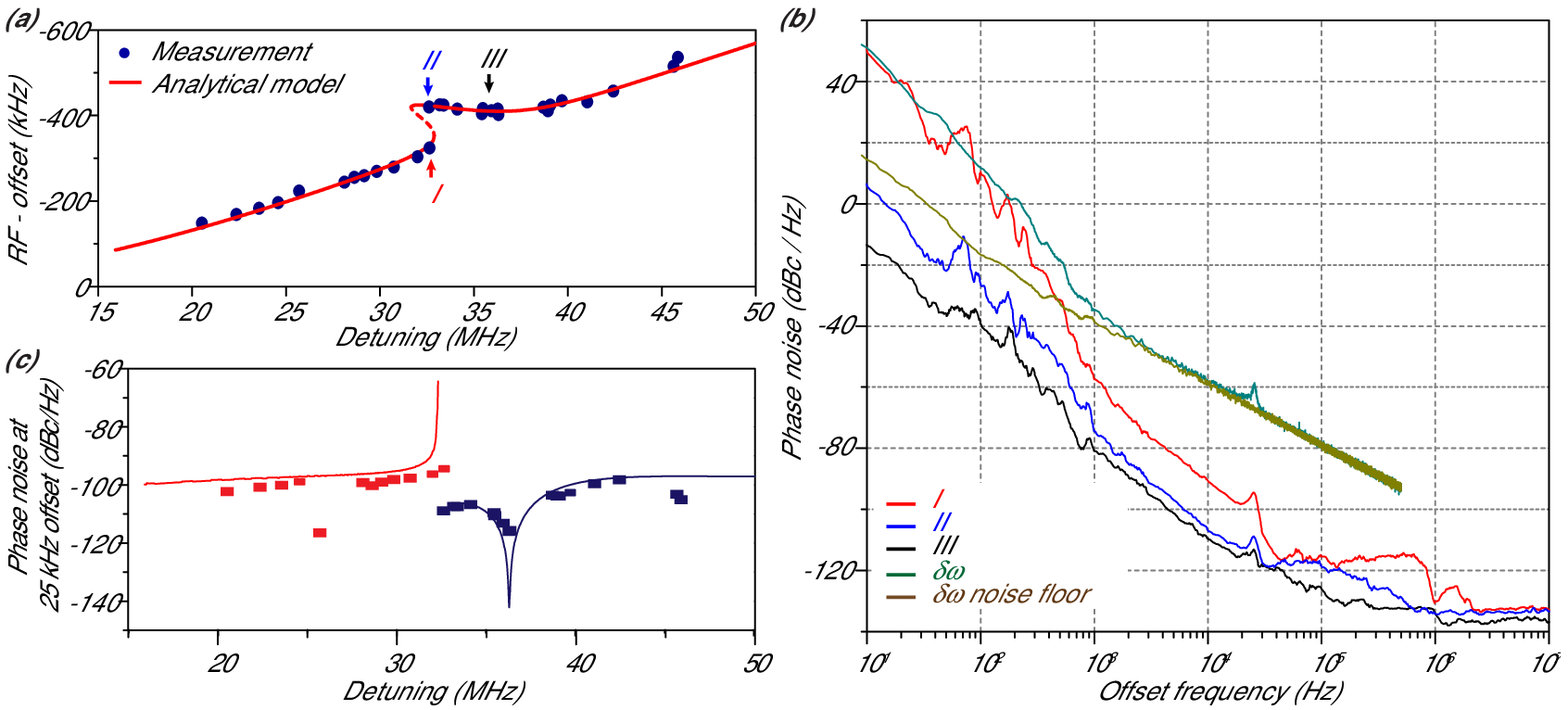}
\caption{{\bf Soliton repetition frequency and phase noise measurement.} {\bf a,} Measured and theoretical soliton repetition frequency versus pump-cavity detuning. The offset frequency is 22.0167 GHz. Operating points I, II and III refer to 3b. Point III is near the quiet operation point. {\bf b,} Phase noise spectra of detected soliton pulse stream at three operating points shown in 3a. {\bf c,} Phase noise of soliton repetition rates at 25 kHz offset frequency plotted versus the cavity-pump detuning. The blue and red dots (lines) denote the experimental (theoretical) phase noise of the upper and lower branch operating points, respectively.}
\label{figure3}
\end{figure*}

\noindent {\bf Experiment and model.} A silica whispering-gallery resonator \cite{lee2012chemically} is used for soliton generation. The devices featured a free-spectral-range (FSR) of approximately 22 GHz (3 mm diameter resonator), intrinsic Q-factors around 250 million and they support multiple, transverse mode families.  In order to characterize the frequency spectrum of the soliton-forming mode family, mode frequencies were measured from 190.95 THz (1570 nm) to 195.94 THz (1530 nm) using an external-cavity diode laser (ECDL) calibrated by a fiber Mach-Zehnder interferometer (MZI) \cite{yi2015soliton}. The measured $relative$-$mode$-$frequency$, $\Delta \omega_{\mu} \equiv \omega_\mu-\omega_o-\mu D_1$ versus mode index, $\mu$, is presented in fig. 1a. In this expression, $\omega_\mu$ is the resonant frequency of the $\mu$-th mode, $D_1$ is the FSR of the soliton-forming mode family at $\mu = 0$ (assumed to be the pumping mode index), and $\omega_0$ is the frequency of mode $\mu = 0$. The dispersion of a mode family can be characterized by expanding the mode frequencies using Taylor series at the pumped mode ($\mu=0$), $\omega_\mu = \omega_0+D_1\mu+\frac{1}{2} D_2\mu^2 + ...$, where $D_2$ is the second order dispersion at $\mu=0$.

DKSs form when the pump laser is red-detuned relative to the pump mode frequency and details on their generation in these resonators are given elsewhere \cite{yi2015soliton,yi2016active}. The soliton-forming mode family (family A in fig. 1a) must have anomalous dispersion near $\mu = 0$ and be relatively free of avoided crossings with other transverse mode families \cite{herr2014mode}. In fig. 1a, the former condition is met by the relative mode frequency spectrum being parabolic with postive curvature near $\mu =0$. A green, dashed parabolic curve ($\Delta \omega_{\mu} = \frac{1}{2} D_2\mu^2$) is provided in the figure to verify this requirement. Concerning the latter condition, a second mode family (family B in fig. 1a) causes an avoided mode crossing near $\mu = 72$ and hybridization of the mode families occurs near the crossing\cite{liu2014investigation,yang2016spatial}. The relative-mode-frequency of the unperturbed soliton-forming mode family and crossing mode family are denoted as $\Delta \omega_{\mu A}$ and $\Delta \omega_{\mu B}$. The lower branch of the hybrid mode family formed by the avoided mode crossing is denoted $\Delta \omega_{\mu -}$. Avoided mode crossing behavior has been intensively studied in the context of DKS solitons. Generally they can interfere with soliton formation \cite{herr2014mode,ramelow2014strong,huang2016smooth}, however, avoided crossings also provide a wave to induce dark soliton formation \cite{xue2015mode}.  In the present system the avoided-mode-crossing induces only minimal distortion in the otherwise parabolic shape of the soliton-forming mode family. Soliton spectra produced on this mode family by pumping at $\mu = 0$ are shown in fig. 1b along with theoretical $\mathrm{sech}^2$ spectral envelopes predicted for DKSs. As an aside, the horizontal scales in fig. 1a and fig. 1b are the same and the location of the $\mu = 0$ pumping mode is indicated by a vertical dashed line in fig. 1b. 

Also shown in fig. 1a are the comb frequencies associated with a fictitious soliton spectrum. This line is given by, 
\begin{equation}
\Delta \omega_{\mu,\mathrm{comb}} = (\omega_{\mathrm{rep}} - D_1) \mu - \delta \omega 
\end{equation}
where $\omega_{\mathrm{rep}}$ is the soliton repetition frequency\cite{yang2016spatial}. (The frequency components of the soliton comb are red-detuned relative to the ``cold-cavity" mode frequencies by the Kerr nonlinearity.) In the vicinity of the avoided crossing it is possible for a spectral component of the soliton, $\mu = \mathrm{r}$ with relative mode frequency $\Delta \omega_{\mathrm{r},\mathrm{comb}}$, to couple resonantly with a mode in the lower branch (relative frequency $\Delta \omega_{\mathrm{r-}}$). The hybrid mode obeys the simple driven oscillator equation,
\begin{equation}
\frac{d h_\mathrm{r-}} {d t} = [-i \Delta \omega_{\mathrm{r-}} - \frac{\kappa_\mathrm{r-}}{2} ] h_{\mathrm{r-}} + f_{\mathrm{r}}e^{-i \Delta \omega_{\mathrm{r, comb}} t }
\end{equation}
where $h_{\mathrm{r-}}$ is the intracavity field amplitude of the mode, $\kappa_\mathrm{r-}$ is its loss rate and $f_{\mathrm{r}}$ is an effective pumping term associated with the soliton comb line at relative frequency $\Delta \omega_{\mathrm{r, comb}}$. The pumping term is given by $f_\mathrm{r}= i\Gamma (\Delta \omega_{\mathrm{r}A}- \Delta \omega_{\mathrm{r,comb}}) a_\mathrm{r}$, where $a_\mathrm{r}$ is the field of the unperturbed soliton hyperbolic solution at $\mu= \mathrm{r}$ (see Methods section). Also, the Kerr effect self-frequency shift of $h_{\mathrm{r-}}$ is of order 10 kHz and is therefore negligible in comparison to $\kappa_{\mathrm{r}-}$. 

Because the lower-branch mode at $\mu = \mathrm{r}$ has a high optical Q factor, slight shifts in the slope of the comb frequency line (equivalently, shifts of $\Delta \omega_{\mathrm{r, comb}}$ relative to $\Delta \omega_{\mathrm{r-}}$) will cause large changes in the power coupled to the mode. These changes are observable in fig. 1b where a strong spectral line appears in the case of the blue soliton spectrum. Note that scattering from the soliton into the spectral line is strong enough so that the power in the line is greater than the comb line power near the spectral center of the soliton, itself. The strong spectral line can be understood as a {\it single-mode dispersive wave} and it induces a recoil in the spectral center of the soliton. This recoil contribution is indicated for the blue soliton spectrum in the figure. In the case of the red soliton spectrum, the operating point was changed and the resonance between the soliton and the mode is diminished. Accordingly, most of the spectral shift in this case results from the Raman SSFS.

A change in the slope of the soliton comb line will occur when the soliton repetition frequency, $\omega_{\mathrm{rep}}$, is changed (see eq. (1)). On account of second order dispersion $\omega_{\mathrm{rep}}$ depends linearly on the frequency offset of the soliton spectral maximum relative to the pump frequency\cite{matsko2013timing,yang2016spatial}. This frequency offset has contributions from both the Raman SSFS, $\Omega_{\mathrm{Raman}}$, and the dispersive-wave recoil, $\Omega_{\mathrm{Recoil}}$ (i.e., $\Omega = \Omega_{\mathrm{Raman}} + \Omega_{\mathrm{Recoil}}$). Accordingly, the soliton repetition rate is given by,
\begin{equation}
\omega_{\mathrm{rep}} =D_1 + \frac{ D_2}{D_1}(\Omega_\mathrm{Raman} + \Omega_\mathrm{Recoil})
\label{eq:Omegarep}
\end{equation}
where $D_2$ is the second order dispersion of soliton-forming mode family at $\mu = 0$ and from fig. 1a is measured to be 17 kHz. Substituting for the repetition rate in the comb line expression (eq. (1)) gives,
\begin{equation}
\Delta \omega_\mathrm{\mu, comb} = \frac{\mu D_2}{D_1} (\Omega_{\mathrm{Raman}} + \Omega_{\mathrm{Recoil}} ) -\delta \omega
\end{equation}
The recoil frequency has a linear dependence on the power of the hybrid mode (see eq.(\ref{eq:Omega_recoil1}) in Methods),
\begin{equation}
\Omega_{\mathrm{Recoil}}=\gamma \left|h_{\mathrm{r-}}\right|^2 = -\mathrm{r}\frac{\kappa_B D_1}{\kappa_A E}\left|h_{\mathrm{r-}}\right|^2,
\label{eq:OmegRec}
\end{equation}
where $\kappa_A$ and $\kappa_B$ denote the loss rates of the family A and family B modes, respectively, and $E$ is the circulating soliton energy. Solving eq. (2) for the steady-state power in the crossing mode at the soliton comb line frequency and using eq. (4)-(5) gives the following,
\begin{equation}
| h_\mathrm{r-} |^2  = \frac{|f_{\mathrm{r}}|^2}{(\Delta \omega_{\mathrm{r-}} + \delta \omega - \frac{\mathrm{r} D_2}{D_1} [\Omega_{\mathrm{Raman}} + \gamma | h_\mathrm{r-} |^2])^2 + \frac{\kappa_\mathrm{r-}^2}{4}}.
\label{eq:hr2}
\end{equation}

Eq. (6) suggests that a hysteresis in the dispersive-wave power is possible when varying the soliton operating point. Consistent with this possibility, it is noted that the two soliton spectra in fig. 1b (blue and red), which show very different dispersive wave powers, were produced at nearly identical detuning frequencies, $\delta \omega$. A more detailed survey of the dispersive wave power behavior is provided in fig. 1c and is again consistent with a hysteretic behavior versus detuning. Also, since the total spectral shift of the soliton is given by $\Omega = \Omega_{\mathrm{Raman}} + \Omega_{\mathrm{Recoil}} = \Omega_{\mathrm{Raman}} + \gamma | h_\mathrm{r-} |^2$ a corresponding hysteresis is observed in the overall soliton spectral shift (fig. 1d). Theoretical fits are provided in fig. 1c and fig. 1d using eq. (6) (see Methods for the fitting procedure and parameter values). In plotting the data, the determination of the detuning frequency, $\delta \omega / 2 \pi$, was made from the measured total soliton spectral shift ($\Omega$) and pulse width ($\tau_s$) using the relation $\delta \omega = D_2/2D_1^2 (1/\tau_s^2+\Omega^2)$ (eq.(\ref{eq:detuningComp}) in Methods)\cite{yi2016theory}.

The recoil frequency, $\Omega_{\mathrm{Recoil}}$, can also be extracted from the data to verify its linear dependence upon dispersive wave power predicted in eq.(\ref{eq:OmegRec}). To do this, the Raman SSFS, $\Omega_{\mathrm{Raman}}$, is determined from the soliton pulse width using eq.(\ref{eq:Omega_Raman1}) in Methods and substracted from the measured total soliton frequency shift. A plot of the recoil shift versus the spur power is given as the inset in fig. 1c and verifies the linear dependence. Eq.(\ref{eq:OmegRec}) is also plotted for comparison using parameters given in the Methods Section. 

While the present results are produced using a dispersive wave that is blue-detuned relative to the soliton spectral maximum, the hysteresis behavior also occurs for a red-detuned dispersive wave. However, in the red-detuned case, the orientation of the curve in fig. 1c is reversed with respect to the detuning frequency. The essential feature for appearance of the hysteresis is that the recoil advances and retreats versus detuning. During a portion of this behavior, it is therefore possible to compensate the Raman SSFS. The requirements imposed on the device and mode crossing for this to occur are discussed below (see section on Existence of quiet point). 

\medskip

\noindent {\bf Numerical simulation.} To attribute the observed frequency shift hysteresis to the single-mode dispersive wave, and to further validate the analytical model, we perform numerical simulations based on the coupled Lugiato-Lefever equations \cite{lugiato1987spatial,matsko2011mode,chembo2013spatiotemporal,d2016nonlinear,matsko2016optical} involving the soliton-forming mode family (family A) and the crossing-mode family (family B). Further information including parameter values is provided in Methods, but is outlined here. The two mode families are coupled using a model studied elsewhere \cite{yang2016spatial}. The coupling is characterized by a rate constant $G$ and is designed to induce an avoided-mode-crossing around mode index $\mu=72$, similar to the experimental mode family dispersion. Fig. 2 shows the results of the numerical simulation including 2048 modes. The hysteresis in the soliton total frequency shift and the dispersive wave power resembles the experimental observation and is also in good agreement with the analytical model (see fig. 2a,b). As predicted by the eq. (\ref{eq:OmegRec}) (and observed in the fig. 1c inset), the recoil is numerically predicted to vary linearly with the dispersive wave power (fig. 2b inset).

Frequency and time domain features of the soliton (blue) and dispersive wave (red) are also studied in fig. 2c,d. They show that the dispersive wave emerges on the crossing-mode family (family B) and consists primarily of a single mode. The single-mode dispersive wave leads to a modulated background field in the resonator with a period determined by the beating between the pump and the dispersive wave. This modulation is observable in fig. 2d. Spectral recoil of the soliton is also observable in the numerical spectra. 

\begin{figure}[!ht]
\captionsetup{singlelinecheck=no, justification = RaggedRight}
\includegraphics[width=\linewidth]{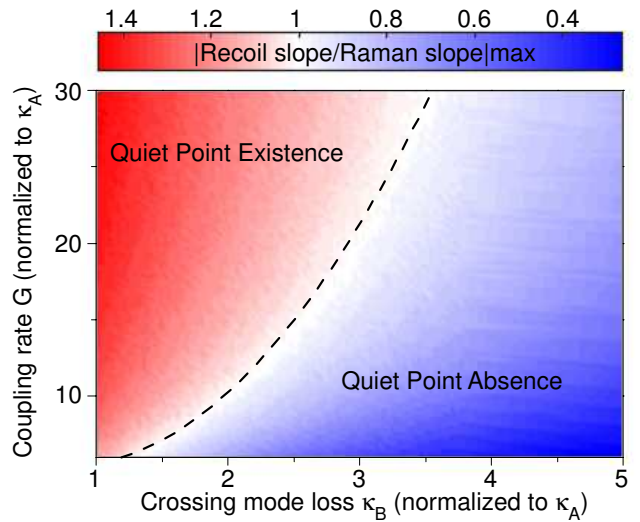}
\caption{{\bf Existence study for the quiet point.} The maximum ratios of $|\partial \Omega_\mathrm{Recoil}/\partial \delta \omega|$ to $|\partial \Omega_\mathrm{Raman}/\partial \delta \omega|$ at varying normalized modal coupling rate $G$ (see Methods) and normalized crossing-mode damping rate $\kappa_B$. The quiet point exists when this ratio is greater than unity (red region).}
\label{figure4}
\end{figure}

\medskip

\noindent {\bf Soliton repetition rate quiet point.} The nonlinear behavior associated with soliton coupling to the single mode dispersive wave can be used to suppress soliton repetition rate noise produced by coupling of pump laser noise. This noise source is suspected to be a significant contributor to repetition rate noise in certain frequency-offset regimes \cite{yi2015soliton}. From eq. (\ref{eq:Omegarep}) the repetition frequency depends linearly on the total soliton spectral-center frequency shift. However, this total shift frequency versus cavity-pump detuning has a stationary point on the upper hysteresis branch (see fig. 1d). As expected from the simple dependence in eq. (\ref{eq:Omegarep}), this same stationary point is observed in measurements of the repetition frequency versus detuning (fig. 3a). To measure the repetition frequency the soliton pulse train is directly detected and an electrical spectrum analyzer is used to observe the pulse train spectrum. The theoretical prediction using analysis from the Methods is also provided for comparison. 

The coupling of pump laser frequency noise into the soliton repetition rate is expected to be minimal at the stationary point. To verify this prediction, the phase noise of the detected soliton pulse train is measured at different soliton operating points on the upper and lower branches in fig. 3a using a phase noise analyzer. Phase noise spectra corresponding to operating points I, II and III in fig. 3a are plotted in fig. 3b. Operating points I and II correspond to nearly identical cavity-pump detuning, but lie on different branches. As expected, operating point II in the upper branch has a lower phase noise level compared to operating point I on account of its reduced slope. Operating point III is close to the zero-slope detuning point in the upper branch. This $quiet$  $point$ has the lowest phase noise among the recorded phase noise spectra. 

For comparison, the phase noise associated with the detuning frequency $\delta \omega$ was also measured. For this measurement, the error signal of a Pound-Drever-Hall feedback control system is operated open-loop and recorded using an oscilloscope. Its power spectral density is converted into phase noise in fig. 3b (see Supplementary Material). The relatively high noise floor in this measurement is caused by the oscilloscope sensitivity. Nonetheless, a noise bump at 25 kHz offset frequency originates from the laser and provides a reference point against which comparison to the soliton phase noise is possible. The soliton phase noise at 25 kHz offset frequency noise is plotted versus detuning in fig. 3c. The calculated soliton phase noise is also presented for comparison using the cavity-pump detunining noise level at 25 kHz offset. The dip of the phase noise occurs at the quiet point. For lower offset frequencies, the contributions to noise are believed to originate from thermal contributions within the resonator and are under investigation. Nonetheless, the measured noise contributions at these frequencies show a trend of reduction for operation at the quiet point.

An analytical study comparing the detuning response of the Raman and recoil effects was performed to determine conditions required to observe the quiet point. The quiet point occurs when the retreating soliton recoil balances the always advancing SSFS. Accordingly, fig. 4 is a contour plot of the maximum ratio of $|\partial \Omega_\mathrm{Recoil}/\partial \delta \omega|$ to $|\partial \Omega_\mathrm{Raman}/\partial \delta \omega|$ while varying the coupling strength between the soliton-mode and crossing-mode families (see Supplementary Material) and the damping rate of the crossing mode. The existence regime for observation of the quiet point corresponds to the ratio $>1$ shown in red. Stronger mode interaction and weaker dissipation are required to operate in this regime. 

\medskip

\noindent{\bf Discussion}

\noindent In summary, soliton coupling to a dispersive wave consisting of a single mode was studied experimentally and theoretically and shown to produce a hysteretic dependence of soliton properties upon cavity-pump detuning. These properties include the frequency shift of soliton spectral center relative to the pumping frequency, the soliton repetition frequency, and the optical power in the single-mode dispersive wave. The accompanying nonlinear response of repetition rate with detuning frequency was shown to create a condition (quiet operating point) where coupling of laser pump frequency noise into the soliton repetition rate is greatly reduced. This reduction was measured by characterizing the soliton pulse-stream phase noise upon photo detection.  The requirements for quiet point existence were also studied. Stronger mode interactions and reduced crossing-mode damping are preferrable. The operating point for quiet soliton operation holds potential for ultra-low-noise microwave generation. 

\medskip

\noindent\textbf{Methods}

\begin{footnotesize}


\medskip

\noindent{\bf Dynamical equation of hybrid mode.}
Equation (2) can be derived from coupled mode equations that include dispersion, mode interaction and the Kerr nonlinearity. The intracavity field of mode $\mu$ in the soliton-forming mode family A can be represented by $A_\mu(t) e^{-i\omega_{\mu A} t+i\mu \phi}$, where $A_{\mu}(t)$ is the slowly varying amplitude, $t$ is the time and $\phi$ is the azimuthal angle along the resonator. In the rotation frame of comb frequencies  $\omega_0 - \delta \omega + \mu \omega_\mathrm{rep}$ for $\mu$, the intracavity field can be expressed as $a_\mu(t) = A_\mu(t) e^{-i(\omega_{\mu A}-\omega_0 + \delta \omega - \mu  \omega_\mathrm{rep})t}$. We denote the intracavity field in the crossing-mode family B as $b_\mu$ and express it in the same reference frame as the soliton-forming mode $a_\mu$. The intracavity fields can be calculated using the equations of motion with Kerr nonlinearity terms \cite{herr2012universal,chembo2013spatiotemporal} and modal-coupling terms\cite{liu2014investigation},
\begin{equation}
\begin{split}
\frac{d a_\mu}{d t} = -\left [\frac{\kappa_A}{2}+i(\omega_{\mu A} - \omega_0 +\delta \omega - \mu \omega_{rep} )\right ] a_\mu + iGb_\mu \\+ig\sum_{j,k}a_j a_k a_{j+k-\mu}^* + F \delta(\mu)
\label{eq:coupleA}
\end{split}
\end{equation}
\begin{equation}
\begin{split}
\frac{d b_\mu}{d t} = -\left [ \frac{\kappa_B}{2}+i(\omega_{\mu B} - \omega_0 +\delta \omega - \mu \omega_{rep} ) \right ] b_\mu+iGa_\mu \\ +ig_B\sum_{j,k}b_j b_k b_{j+k-\mu}^* 
\label{eq:coupleB}
\end{split}
\end{equation}
where $\kappa_{A,B}=\omega_{0}/Q_{A,B}$ is the dissipation rate;
$g=\hbar\omega_{0}^2 n_{2}D_{1}/2\pi n_0 A_{eff}$ represents the normalized Kerr nonlinear coefficient where $A_{eff}$ is the effective nonlinear mode area and $g_B$ is defined in the same way. $G$ is the linear coupling coefficient between the two mode families \cite{yang2016spatial} and $F$ is the coupled laser pump field. Also, to calculate eq. (2) it is not necessary to include Raman coupling terms in eq.(\ref{eq:coupleA}) and eq.(\ref{eq:coupleB}) since the leading-order contribution to the forcing term, $f_r$,  is from the Kerr nonlinearity. 


Modal coupling forms two branches of hybrid modes measured in the mode spectrum (fig.1a). The frequency of the hybrid modes in the upper (+) and lower (-) branches is given by \cite{haus1991coupled,wiersig2006formation,liu2014investigation}
\begin{equation}
\omega_{\mu \pm} = \frac{\omega_{\mu A}+\omega_{\mu B}}{2} \pm \sqrt{G^2+\frac{1}{4}(\omega_{\mu A}-\omega_{\mu B})^2}
\end{equation}
where the corresponding field amplitude of the hybrid modes is a linear combination of $a_\mu $ and $b_\mu$.  In the far-detuned regime where $\omega_{\mu A} - \omega_{\mu B} \gg G$, the field amplitude of the lower branch hybrid mode is approximately given by, 
\begin{equation}
\widetilde{h}_{\mu -} = \frac{Ga_\mu + (\omega_{\mu A}-\omega_{\mu B })b_\mu}{\sqrt{G^2+(\omega_{\mu A}-\omega_{\mu B})^2}}.
\label{eq:hybCombi}
\end{equation}

\noindent In this experiment, only one mode was observed to be near resonance with the soliton comb and that mode is assigned mode index $\mu = r$. Also, as the amplitude of $b_\mu$ with $\mu \neq r$ is small, the Kerr interaction summation term can be neglected in eq.(\ref{eq:coupleB}) in this calculation. 

By taking the time derivative of eq. (\ref{eq:hybCombi}) and then substituting using (\ref{eq:coupleA}) and (\ref{eq:coupleB}) the following dyanamical equation results for $\widetilde{h}_{\mu \pm}$,
\begin{equation}
\frac{d \widetilde{h}_{r -} }{d t} = -\left [ \frac{\kappa_{r-}}{2} + i(\omega_{r-}  - \omega_0 +\delta \omega - \mathrm{r} \omega_{rep} ) \right] \widetilde{h}_{r -}+f_{r} 
\label{eq:hmur_}
\end{equation}
where $f_{r}$ is the pumping term given by,
\begin{equation}
f_{r} = i \Gamma g\sum_{j,k}a_j a_k a_{j+k-r}^*.
\label{eq:feff}
\end{equation}
and where $\Gamma=G/\sqrt{|G|^2+|\omega_{\mu A}-\omega_{\mu B}|^2}$ is the weight of the family A mode in  $\widetilde{h}_{\mu-}$ and $\kappa_{r-} \approx \kappa_B$ for $r$ when $\Gamma \ll 1$. Also, consistent with fig. 1a, the hybridization of mode r is assumed weak (i.e., $|\omega_{r A}-\omega_{r B}|\gg |G|$ and $ |\Delta \omega_{r A}| \gg |\Delta \omega_{r B}|$) so that $b_r$ is the dominant contribution to $\widetilde{h}_{r -}$. When converting eq. (\ref{eq:hmur_}) into the rotation frame of $(\omega_0+\mu D_1)$ with $\widetilde{h}_{r-} = h_{r-}e^{i\Delta \omega_{r,\mathrm{comb}}t}$, the following expression results,
\begin{equation}
\frac{d h_{r-}}{d t} = [-i\Delta \omega_{r-}-\frac{\kappa_{r-}}{2}]h_{r -}+f_{r}e^{-i\Delta \omega_{r,\mathrm{comb}}t}
\label{eq:Maineq1}
\end{equation}
where $\Delta \omega_{r -} = \omega_{r-} - \omega_o - \mu D_1 $ is the relative-mode-frequency of hybrid mode ${h}_{r-}$. {\it Equation (\ref{eq:Maineq1}) is identical to eq.(2) in the main text.}
 
\medskip


\medskip

\noindent{\bf Effective pumping term.}
The pumping term in eq.(\ref{eq:hmur_}) can be expressed in parameters of the resonator and soliton. The soliton field envelope takes the form \cite{herr2014temporal,yi2016theory}
\begin{equation}
A(\phi,t) = B_s\mathrm{sech}[(\phi-\phi_c)/D_1\tau_s]e^{i\Omega(\phi-\phi_c)/D_1+i\varphi}
\label{eq:ansatzLag}
\end{equation}
where soliton properties are: amplitude $B_s$, angular position $\phi_c$, temporal pulse width $\tau_s$, spectral-center frequency shift (relative to pump) $\Omega$ and phase relative to the pump laser $\varphi$ . Also, this solution assumes $\delta \omega \gg \kappa_A$. By applying the Fourier transform to $A(\phi,t)$, $a_\mu$ can be expressed in terms of the soliton properties,
\begin{equation}
A (\phi, t) = \sum_\mu a_\mu (t) e^{i\mu (\phi-\phi_c)},
\label{eq:Aphit}
\end{equation}
\begin{equation}
a_\mu = \frac{B_s\tau_sD_1}{2} \mathrm{sech}(\frac{\pi\tau_s }{2}(D_1\mu-\Omega))e^{i\varphi}.
\label{eq:def_amu}
\end{equation}

\noindent  The pump $f_r$ can therefore be derived by inserting eq.(\ref{eq:def_amu}) into eq.(\ref{eq:feff}). The following expression results from simplification of the summation, 
\begin{equation}
f_{r} = i\Gamma \frac{D_2} {4 D_1^2}[(D_1 r-\Omega)^2+\frac{1}{\tau_s^2}] B_s\tau_s D_1 \mathrm{sech}(\frac{\pi\tau_s }{2}(D_1r-\Omega))e^{i\varphi}
\label{eq:feff_comp}
\end{equation}
where $g$ has been replaced using equation $B_s^2\tau_s^2 = D_2/gD_1^2$, which holds for DKSs \cite{matsko2013timing,yi2016theory} and is also verified in a section below. Finally, by using \cite{yi2016theory} $\delta \omega = \frac{D_2}{2D_1^2}(\frac{1}{\tau_s^2}+\Omega^2)$ (see derivation below), $f_r$ can be further reduced to 
\begin{equation}
f_{r} = i\Gamma (\Delta \omega_{rA}- \Delta \omega_{r,\mathrm{comb}}) a_r.
\label{eq:feff_comp2}
\end{equation}

\medskip

\noindent{\bf Recoil and Soliton Self Frequency Shift.}
In addition to the Raman SSFS \cite{karpov2016raman,yi2016theory}, the spectral center of the DKS can also be shifted by the single line dispersive wave recoil. The effect of the recoil and Raman shift can be calculated using the moment analysis method \cite{santhanam2003raman,karpov2016raman}. 
Using the Fourier transform, eq.(\ref{eq:coupleA}) is transformed into the perturbed Lugiato-Lefever equation (LLE) \cite{chembo2013spatiotemporal}
\begin{equation}
\begin{split}
\frac{\partial A(\phi, t)}{\partial t}= -(\frac{\kappa_A}{2} +i\delta \omega)A+i\frac{D_2}{2}\frac{\partial^2 A}{\partial \phi^2}+F
+ig|A|^2A  \\
+ig\tau_{R}D_1A\frac{\partial |A|^2}{\partial \phi}+iGB,
\label{eq:LLEA}
\end{split}
\end{equation}
where the Raman shock term has been added \cite{karpov2016raman,yi2016theory} and $\tau_{R}$ is the Raman time constant. The moment analysis method treats the soliton as a particle. The energy E and the spectral center mode number $\mu_c$ are given by,
\begin{equation}
E = \sum_\mu|a_\mu|^2 = \frac{1}{2\pi}\int_{-\pi}^{+\pi}|A|^2\mathrm{d}\phi = B_s^2\tau_sD_1/\pi
\label{eq:E}
\end{equation}
\begin{equation}
\mu_c = \frac{\sum_\mu\mu|a_\mu|^2}{E} = \frac{-i}{4\pi E}\int_{-\pi}^{+\pi}(A^*\frac{\partial A}{\partial \phi}-A\frac{\partial A^*}{\partial \phi})\mathrm{d}\phi.
\label{eq:muc}
\end{equation}
Taking the time derivative of eq.(\ref{eq:muc}) and substituting $\partial A/ \partial t$ using eq.(\ref{eq:LLEA}), the following equation of motion for $\mu_c$ is obtained,
\begin{equation}
\begin{split}
\frac{\partial \mu_c}{\partial t} = -\kappa_A \mu_c -\frac{g\tau_{R}D_1}{2\pi E}\int_{-\pi}^{+\pi}(\frac{\partial |A|^2}{\partial \phi})^2 \mathrm{d} \phi \\
-\frac{1}{2\pi E} \int_{-\pi}^{+\pi} (G^* B^*\frac{\partial A}{\partial \phi}-GA^*\frac{\partial B}{\partial \phi})\mathrm{d}\phi.
\label{eq:pmuc}
\end{split}
\end{equation}
The second term on the right-hand-side corresponds to the Raman-induced frequency shift and the third term is the frequency shift caused by recoil. 

The Raman term can be calculated by substituting eq.(\ref{eq:ansatzLag}) into the integral. When calculating the recoil term, $B$ is simplified to $B \approx b_{r}e^{ir (\phi-\phi_c)}$ as the power in mode $B$ is dominated by the near resonance mode $r$. In addition, because the integral of $\phi$ is over $2\pi$, only $a_r e^{ir (\phi-\phi_c)}$ has nonzero contribution. Furthermore, equation (\ref{eq:coupleB}) is used to relate $G a_r$ to $b_r$ and finally leads to,
\begin{equation}
\frac{\partial \mu_c}{\partial t} = -\frac{8\tau_{R}D_2}{15D_1^3\tau_s^4}-\frac{r\kappa_B}{E}|b_{r}|^2-\kappa_A \mu_c,
\end{equation}
The steady-state spectral center mode number is therefore given by, 
\begin{equation}
\begin{split}
\mu_c = -\frac{8\tau_{R}D_2}{15\kappa_AD_1^3\tau_s^4}-\frac{r\kappa_B}{\kappa_AE(1-\Gamma^2)}|h_{r-}|^2 \\
= \frac{1}{D_1}(\Omega_{\mathrm{Raman}}+\Omega_{\mathrm{Recoil}}),
\label{eq:mucfn1}
\end{split}
\end{equation}
where $|\omega_{\mu A}-\omega_{\mu B}| \gg \kappa_B, \Delta \omega_{r-}$ (equivalent to $|b_r| \gg |a_r|$) is assumed and the recoil and Raman shifts are,
\begin{equation}
\Omega_{\mathrm{Recoil}} = \gamma|h_{r-}|^2 = -\frac{r\kappa_BD_1}{\kappa_AE(1-\Gamma^2)}|h_{r-}|^2,
\label{eq:Omega_recoil1}
\end{equation}
\begin{equation}
\Omega_{\mathrm{Raman}} = -\frac{8\tau_{R}D_2}{15\kappa_AD_1^3\tau_s^4}
\label{eq:Omega_Raman1}
\end{equation}
where in the main text, $\Gamma^2 \ll 1 $ is assumed. {\it Eq.(\ref{eq:Omega_recoil1}) is eq.(\ref{eq:OmegRec}) in the main text.} The form for the Raman SSFS, $\Omega_{\mathrm{Raman}}$, is identical to the form previously derived in the absence of the dispersive-wave coupling \cite{yi2016theory}.



\medskip

\noindent {\bf Soliton parameters with Raman and mode-coupling effects.} In the presence of recoil and Raman, the relations between soliton parameters in eq.(\ref{eq:ansatzLag}) can be derived from the Lagrangian approach \cite{yi2016theory,matsko2013timing,herr2014temporal}. In addition, the Lagrangian approach verifies the expression for $\Omega_{\mathrm{Recoil}}$ obtained above as well as providing a path for calculation of the repetition-rate phase noise \cite{matsko2013timing}. As detailed in previous literature\cite{matsko2013timing,yi2016theory}, the perturbation Lagrangian method is applied to the LLE equation of $A$ (eq. \ref{eq:LLEA}). However, now an additional perturbation term is added to account for the mode coupling to the crossing-mode family. Taking $B \approx b_{r}e^{ir (\phi-\phi_c)}$, produces the following equations of motion,

\begin{equation}
\frac{\Omega}{D_1}  \frac{\partial{\phi_c}}{\partial{t}}-\frac{\partial{\varphi}}{\partial{t}}-\delta \omega-\frac{D_2 \Omega^2}{2D_1^2}-\frac{D_2}{6\tau_s^2D_1^2} +\frac{2}{3}gB_s^2 = 0
\label{eq:motB}
\end{equation}
\begin{equation}
\frac{\Omega}{D_1} \frac{\partial{\phi_c}}{\partial{t}}-\frac{\partial{\varphi}}{\partial{t}}-\delta \omega-\frac{D_2 \Omega^2}{2D_1^2}+\frac{D_2}{6\tau_s^2D_1^2} +\frac{1}{3}gB_s^2 =0
\label{eq:mottaus}
\end{equation}
\begin{equation}
	\frac{\partial{(B_s^2\tau_s\Omega)}}{\partial{t}} = -\kappa_A B_s^2\tau_s\Omega-\frac{8g\tau_RB_s^4}{15\tau_s}-\kappa_B\pi r|b_r|^2
\label{eq:mott0}
\end{equation}
\begin{equation}
\frac{\partial{\phi_c}}{\partial{t}} = \frac{D_2}{D_1}\Omega
\label{eq:motOmg}
\end{equation}
\begin{equation}
	\frac{\partial{(B_s^2\tau_s)}}{\partial{t}} = -\kappa_A B_s^2\tau_s+f\cos \varphi B_s \tau_s \pi \mathrm{sech} (\Omega \tau_s \frac{\pi}{2})
	\label{eq:motph}
\end{equation}
where we have assumed the mode r is far from the mode center $\mu_c=\Omega/D_1$ and the coupling coefficient $G$ is smaller than or around the same order of magnitude with $\delta \omega$. Also, higher order terms are neglected (see Supplement).  Subtracting eq.(\ref{eq:mottaus}) from eq.(\ref{eq:motB}) yields 
\begin{equation}
B_s\tau_s = \sqrt{\frac{D_2}{gD_1^2}}
\label{eq:Btaus}
\end{equation}
This equation was previously verified in the presence of Raman-only interactions \cite{yi2016theory}. 

An additional relation between $\delta \omega$, $\tau_s$ and $\Omega$ is derived for steady state by substituting eq.(\ref{eq:motOmg}) and (\ref{eq:Btaus}) into eq.(\ref{eq:motB})：

\begin{equation}
\delta \omega = \frac{D_2}{2D_1^2}(\frac{1}{\tau_s^2}+\Omega^2).
\label{eq:detuningComp}。
\end{equation}
where $\Omega$ can be obtained from (\ref{eq:mott0}) and (\ref{eq:Btaus}),
\begin{equation}
\Omega = \Omega_{\mathrm{Raman}}+\Omega_{\mathrm{Recoil}} = -\frac{8D_2\tau_R}{15\kappa_AD_1^2\tau_s^4}-\frac{r\kappa_BD_1}{\kappa_AE(1-\Gamma^2)}|h_{r-}|^2
\label{eq:LagOmeg}
\end{equation}
which provides an independent confirmation of eq.(\ref{eq:mucfn1}). Also, eq. (\ref{eq:detuningComp}) is identical in form to an expression which included only the Raman SSFS\cite{yi2016theory}. Significantly, however, eq. (\ref{eq:detuningComp}) is more general since $\Omega$ is the total spectral center shift provided by the combined effects of Raman SSFS and dispersive-wave recoil. 

\medskip

\medskip
\noindent{\bf Analytical model fitting and parameters.} Measurements are compared with the analytical model in figures 1c, 1d and 3a. Measured parameters used for the analytical model are: $\kappa_A/2\pi=2.12$ MHz, $D_1/2\pi=22$ GHz, $D_2/2\pi=17$ kHz, $G/2\pi=42.4$ MHz. $\tau_R=2.49$ fs can be extracted from measured $\Omega$ in the regime without the mode recoil effect ($\delta \omega/2\pi < 30$ MHz and $\delta \omega/2\pi > 40$ MHz). Two free parameters are used to optimize the fitting in figure 1 and 3 and they are in reasonable agreement with the measurement: $\Delta\omega_\mathrm{r-}=-62.2$ MHz ($-75\pm 7$ MHz in measurement) and $\kappa_{r-}/2\pi=3.6$ MHz ($6$ MHz in measurement). The procedure for fitting is as follows: a detuning frequency, $\delta \omega$, (horizontal axis in fig. 1c, 1d and 3a plots) is selected. By eliminating $\Omega$ in eq. (\ref{eq:detuningComp}) and eq. (\ref{eq:LagOmeg}) a single condition relating $\tau_s$ and $|h_{r-}|^2$ results. Likewise, with $\delta \omega$ selected a second condition relating $\tau_s$ and $|h_{r-}|^2$ results from eq. (\ref{eq:hr2}) by replacing $\Omega_{\mathrm{Raman}}$ using eq. (\ref{eq:Omega_Raman1}). This pair of equations is solved for $\tau_s$ and $|h_{r-}|^2$ from which $\Omega$ is determined by eq. (\ref{eq:LagOmeg}) and $\omega_{\mathrm{rep}}$ is determined by eq. (\ref{eq:Omegarep}).

\medskip

\noindent{\bf Numerical Simulations.} Numerical simulations based on the coupled Lugiato-Lefever equation of mode family A and B (eq. (\ref{eq:LLEA}) and Fourier transform of eq. (\ref{eq:coupleB})) are implemented to further validate the analytical model. The Raman term in mode family B is ignored since the power in mode family B is too small to induce Raman related effects. Dispersion of 3rd order and higher as well as the self-steepening effect \cite{agrawal2007nonlinear} are neglected. The simulations are implemented with the split-step Fourier method \cite{agrawal2007nonlinear} where 2048 modes in the frequency domain are taken into account. The parameters for two mode families used in figure 2 and 4 are $\kappa_{A}/2\pi = $2.12 MHz, $\kappa_B/2\pi = 3.4$MHz, $D_1/2\pi = $22 GHz for mode A, $D_{1B}/2\pi = D_1/2\pi + 50.9$ MHz for mode B, $D_2/2\pi = $17 kHz for both mode A and B, $\tau_{R} = $2.489 fs, $g = g_{B} = 9.8\times10^{-4}$rad/s and $G/2\pi = $42.4 MHz. 

\end{footnotesize}

\medskip

\noindent\textbf{Acknowledgment}

\noindent The authors gratefully acknowledge the Defense Advanced Research Projects Agency under the QuASAR and PULSE programs, the Kavli Nanoscience Institute. Xueyue Zhang gratefully thanks the Caltech SURF program and the Tsinghua University Top Open Program, Spark Program and Initiative Scientific Research Program (No.20161080166).

\bibliography{ref}

\pagebreak
\widetext
\begin{center}
\textbf{\large Supplemental Information: Single-mode dispersive waves and soliton microcomb dynamics}
\end{center}
\setcounter{equation}{0}
\setcounter{figure}{0}
\setcounter{table}{0}
\makeatletter
\renewcommand{\theequation}{S\arabic{equation}}
\renewcommand{\thefigure}{S\arabic{figure}}
\renewcommand{\bibnumfmt}[1]{[S#1]}
\renewcommand{\citenumfont}[1]{S#1}
\begin{center}
Xu Yi$^{1,\ast}$, Qi-Fan Yang$^{1,\ast}$, Xueyue Zhang$^{1,2,\ast}$, Ki Youl Yang$^{1}$, and Kerry Vahala$^{1,\dagger}$\\
$^{1}$T. J. Watson Laboratory of Applied Physics, California Institute of Technology, Pasadena, California 91125, USA.\\
$^{2}$Department of Microelectronics and Nanoelectronics, Tsinghua University, Beijing 100084, P. R. China\\
$^{\ast}$These authors contributed equally to this work.\\
$^{\dagger}$Corresponding author: vahala@caltech.edu
\end{center}

\section{Phase noise transfer function}

The repetition rate of the soliton can be expressed as follows \cite{syang2016spatial},
\begin{equation}
\omega_{\mathrm{rep}} =  D_1+\frac{\partial \phi_c}{\partial t} = D_1 +\frac{D_2}{D_1}\Omega
\label{eq:solitonrep}
\end{equation}
The variation in both $D_1$ and $\Omega$ contribute to fluctuations in the repetition rate. While $D_1$ is subject to thermo-refractive noise and fluctuations from the environment, a significant contributor to fluctuations in $\Omega$ result from the pump-laser frequency detuning noise (as shown in main text $\Omega$ is a function of cavity-laser detuning). This noise can be calculated using the Lagrangian approach\cite{smatsko2013timing} and is given below. In the following derivation, $\tau_s$ in eq.(27)-(31) in the Methods section is eliminated using eq.(32).

Small-signal noise contributions associated with the frequency detuning and pump-laser power are denoted as $\widetilde{\delta \omega (\omega)}$ and $\widetilde{f(\omega)}$, where $\omega$ is the phase-noise offset frequency. These noise contributions are coupled to soliton parameters through eq.(27)-(31) in the Methods section. By applying the Fourier transform to the eq.(27)-(31) and substituting the pump laser noise contributions $\widetilde{\delta \omega (\omega)}$ and $\widetilde{f(\omega)}$, the resulting small-signal variation of all soliton parameters can be computed. Specifically, it can be shown,

\begin{equation}
\begin{split}
\widetilde{\frac{\partial \phi_c (\omega)}{\partial t}} =\frac{D_2}{D_1} \widetilde{\Omega} = \frac{D_2}{ D_1} (\tan \varphi \widetilde{ \delta \omega}	(\omega)+ \frac{i\omega \widetilde{ f(\omega)}}{f})\\ \times \frac{\kappa_A B_{s}C_1(\omega)}{-i\omega B_{s}C_2(\omega)-\kappa_A B_{s} C_2(\omega) -C_1(\omega)C_3(\omega)} 
\end{split}
\label{eq:vari}
\end{equation}
where sources of noise associated with $D_1$ in eq. (\ref{eq:solitonrep}) are ignored. Also, the following quantities are defined,  
\begin{equation}
C_1(\omega) = i\omega \Omega - \kappa_A B_{s} \frac{\partial \delta \omega}{\partial B_s} \frac{\partial \Omega}{\partial \delta \omega}
\label{eq:C1w}
\end{equation}

\begin{equation}
\begin{split}
C_2(\omega) = -\omega^2+i\omega\kappa_A-i\omega\kappa_A B_{s} \mathrm{tanh}(\sqrt{\frac{D_2}{g}}\frac{\pi\Omega}{2B_{s}D_1}) \\
\sqrt{\frac{D_2}{g}}\frac{\pi\Omega}{2B_{s}^2D_1}
+\kappa_A g B_{s}^2\tan \varphi 
\end{split}
\end{equation}
\begin{equation}
\begin{split}
C_3(\omega) = -i\omega\kappa_AB_{s}\mathrm{tanh}(\sqrt{\frac{D_2}{g}}\frac{\pi\Omega}{2B_{s}D_1})\sqrt{\frac{D_2}{g}}\frac{\pi}{2B_{s}D_1} \\
-\frac{D_2}{D_1^2}\kappa_AB_{s} \Omega \tan \varphi
\end{split}
\end{equation}

\noindent When the repetition rate noise is dominated by the detuning noise term,  the soliton phase noise can be expressed as $\widetilde{\omega_{\mathrm{rep}}}(\omega) = \alpha(\omega) \widetilde{\delta \omega}	(\omega)$, where $\alpha(\omega)$ is the coefficient of $\widetilde{\delta \omega}	(\omega)$ in eq. S2. It is, in effect, the noise transfer function.  Accordingly, the phase noise of repetition rate is $S_\phi (\delta \omega) = |\alpha(\omega)|^2 S_{\phi,\delta \omega} (\omega)$. When $\partial \Omega/ \partial \delta \omega$ appoaches zero, the second term in $C_1$ vanishes and noise transfer function reaches a minimum.

\textit{The result of the converted phase noise is shown by the curves in fig.3c (main text).}  Here $\Omega_{\mathrm{Raman}}$ and $\Omega_{\mathrm{Recoil}}$ are extracted numerically from the fitting curves in fig.1  using the expressions in the previous sections. 

\section{Experimental Setup and detuning noise measurement}

\begin{figure}[!ht]
\captionsetup{singlelinecheck=no, justification = RaggedRight}
\includegraphics[width=\linewidth]{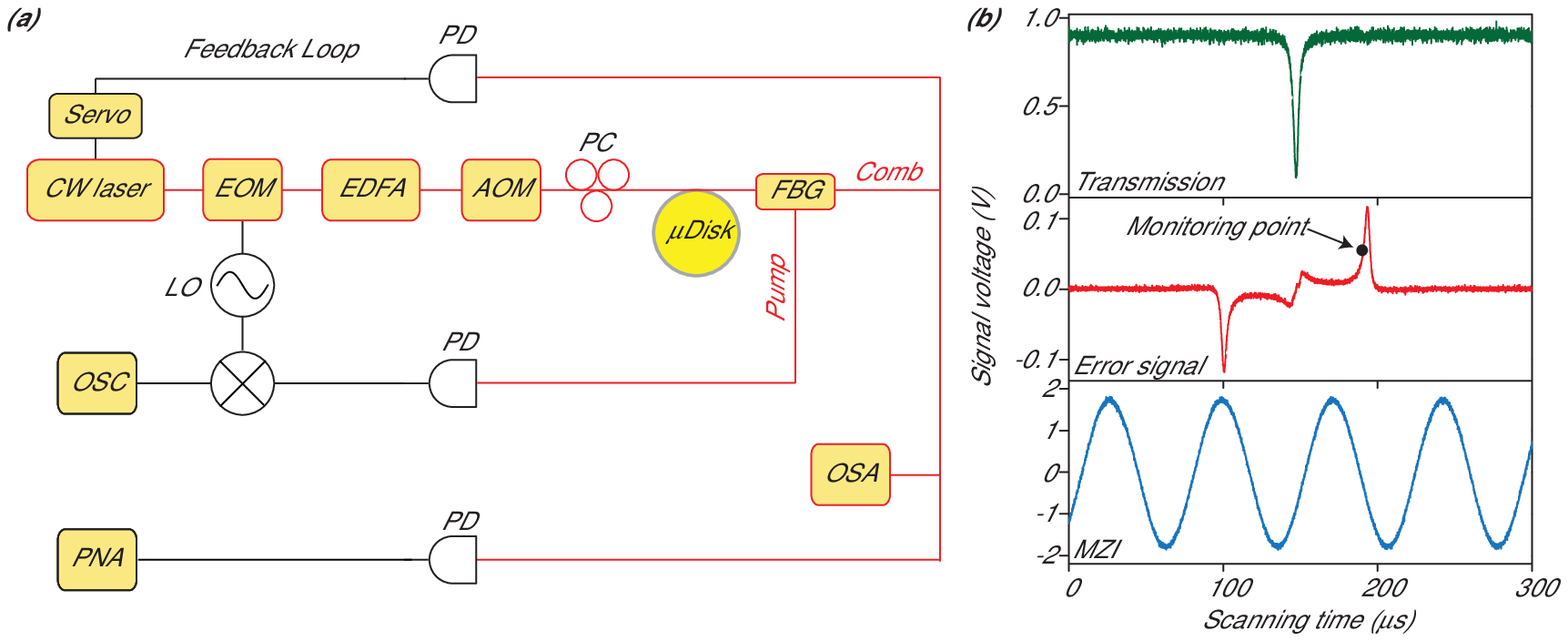}
\caption{Experimental setup and details on detuning-noise measurement. (a) The experimental setup includes both the soliton generation and characterization setup \cite{syi2015soliton,syi2016active} and a Pound-Drever-Hall (PDH) system operated open loop. The PDH is added to make possible the pump-cavity detuning noise measurement. Components included in the set up are an EOM: electro-optic modulator; EDFA: Erbium-doped fiber amplifier; AOM: acousto-optic modulator; PC: polarization controller; FBG: fiber Bragg grating; PD: photodetector; OSA: optical spectral analyzer; PNA: phase noise analyzer; LO: local oscillator. The OSA and and PNA are shown for completeness. They are used to measure the soliton spectrum and repetition rate phase noise. They are not involved in measuring the detuning frequency noise. (b) Measurements that illustrate the pump-cavity detuning measurement. Upper panel (green) shows the measured power transmission when scanning the pump laser frequency across a cavity resonance. In the middle panel, the pump laser is phase modulated, the transmitted signal is detected and the resulting photocurrent is then mixed with the PDH local oscillator signal to generate the PDH error signal. Upon laser scan the PDH error signal (as measured on the oscilloscope) is generated as shown in the middle panel (red). The pump laser is filtered using the fiber Bragg grating. The monitoring point for the detuning frequency measurement is indicated in the middle panel. In order to convert scanning time in the upper and middle panels into laser frequency, a calibrated Mach-Zehnder interferometer (MZI) records power transmission (blue) on an oscilloscope. The free-spectral-range of the MZI is 40MHz.}
\label{figsuppdh}
\end{figure}

Solitons are generated and locked using the active-capture and locking technique \cite{syi2015soliton,syi2016active}. In this method a feedback loop controls the pump laser frequency (fiber laser is used in this work) to maintain soliton power. In order to measure the detuning noise, an additional Pound-Drever-Hall (PDH) loop is embedded into the setup and operated open loop (see Fig. \ref{figsuppdh}a). The pump frequency is red detuned relative to the cavity resonance in order to form the soliton pulse train. Moreover, the amount of cavity-laser detuning required to generate solitons is many cavity linedwidths so that the conventional PDH error signal near the resonance frequency cannot be used to monitor the detuning frequency. However, the higher-frequency PDH sideband can be tuned to reside close to the cavity resonance. Path phases in the PDH loop can be adjusted so that a PDH error signal is produced by the interaction of this sideband with the cavity resonance. When the soliton is formed, we tune the PDH local-oscillator (LO) frequency to approximately match the cavity-laser detuning. This is accomplished by monitoring the PDH error signal (see red trace in middle panel of fig. S1b). For this measurement the transmitted pump light is filtered from the soliton spectrum using a fiber Bragg filter. By setting this LO frequency to the indicated monitoring point, the corresponding error-signal output will convert detuning frequency to a voltage output. This output can be recorded and then analyzed to produce a noise spectrum. The calibration of voltage into frequency is performed by using the Mach-Zehnder interferometer trace (see blue trace in fig. S1b). This calibration is performed on the resonator at reduced power levels where solitons do not form and where the Lorentzian lineshape of the resonator is unaffected by the Kerr nonlinearity. 

\section{Approximations in the equations of motion}

The coupling with mode family B results in the recoil term in eq.(29) in the Methods section. In eq. (27), (28), (30) and (31) (Methods section), higher order terms have been neglected. Here, we list the higher order terms versus the leading order terms in each equation and therefore establish the validity conditions for these equations. 
\begin{equation}
\frac{2\pi|G|^2\tau_sD_1 \Delta \omega_{\mathrm{r}B}' \mathrm{sech}^2(\pi\tau_s(rD_1-\Omega)/2)}{gB_s^2 (\kappa_B^2+4\Delta \omega_{\mathrm{r}B}'^2)} \le \frac{\pi |G |^2 | a_r |^2}{\kappa_B \delta \omega B_s^2 \tau_s D_1} \ll 1
\label{eq:s1}
\end{equation}
\begin{equation}
\frac{2\pi|G|^2\tau_sD_1\Delta \omega_{\mathrm{r}B}' \mathrm{sech}^2(\pi\tau_s(rD_1-\Omega)/2)[2-\Omega\tau_s\pi\mathrm{tanh}(\Omega\tau_s\pi/2)]}{gB_s^2 (\kappa_B^2+4\Delta \omega_{\mathrm{r}B}'^2)} \le \frac{2\pi |G |^2 | a_r |^2}{\kappa_B \delta \omega B_s^2 \tau_s D_1} \ll 1
\end{equation}
\begin{equation}
\frac{|G|^2D_1\mathrm{sech}(\pi\tau_s(rD_1-\Omega)/2)\mathrm{F}(\Omega\tau_s)}{2\kappa_B gB_s^2\Omega} \approx \frac{|G|^2|a_r|}{2\kappa_B\delta \omega \Omega} \ll 1
\label{eq:wOmeg}
\end{equation}
\begin{equation}
\frac{\pi|G^2|B_s^2\tau_s^2D_1\kappa_B\mathrm{sech}^2(\pi\tau_s(rD_1-\Omega)/2)}{\kappa_AB_s^2\tau_s (\kappa_B^2+4\Delta \omega_{\mathrm{r}B}'^2)} \le \frac{4\pi|G|^2|a_r|^2}{\kappa_A\kappa_BB_s^2\tau_sD_1} \ll 1
\label{eq:s4}
\end{equation}
where $\mathrm{F}(\Omega\tau_s)=\frac{i\pi^2}{8} [\csc^2(\frac{1+\Omega \tau_s i}{4}\pi)+\mathrm{csch}^2(\frac{\Omega \tau_s+i}{4}\pi)]\sim O(1) $ in eq.(\ref{eq:wOmeg}) and $\Delta \omega_{\mathrm{r}B}' = \Delta \omega_{\mathrm{r}B} - \Delta \omega_{\mathrm{r,comb}}$. Eq.(\ref{eq:s1})-(\ref{eq:s4}) hold under the experimental conditions and ignoring these terms gives the equations of motion eq.(27)-(31) in the Methods section.

\end{document}